\documentclass[
    ,final            
  ]
  {aipproc}
\layoutstyle{8x11double}
\usepackage{enumerate,amsmath,xspace}
\newcommand{\taus}{\ensuremath{\tau_\text{s}\xspace}}
\newcommand{\taueq}{\ensuremath{\tau_\text{eq}\xspace}}

\begin{document}
\title{Flow instabilities in complex fluids: \\Nonlinear rheology and slow relaxations}
\author{A.~Aradian}{
  address={School of Physics, University of Edinburgh, JCMB Kings Buildings, Edinburgh EH9 3JZ, United
  Kingdom \emph{E-mails:} A.Aradian@ed.ac.uk, M.E.Cates@ed.ac.uk}
}
\author{M.~E.~Cates}{
  address={School of Physics, University of Edinburgh, JCMB Kings Buildings, Edinburgh EH9 3JZ, United
  Kingdom \emph{E-mails:} A.Aradian@ed.ac.uk, M.E.Cates@ed.ac.uk}}
\begin{abstract}
We here present two simplified models aimed at describing the
long-term, irregular behaviours observed in the rheological
response of certain complex fluids, such as periodic oscillations
or chaotic-like variations. Both models exploit the idea of having
a (non-linear) rheological equation, controlling the temporal
evolution of the stress, where one of the participating variables
(a ``structural'' variable) is subject to a distinct dynamics with
a different relaxation time. The coupling between the two dynamics
is a source of instability.
\end{abstract}
\maketitle
%
%
%
\subsection{Introduction}
Complex fluids are known to exhibit a wide range of unconventional
behaviour when forced to flow, due to the intricate couplings
between their structure at the mesoscopic scale and the imposed
flow. In this work, we investigate theoretically certain
situations where, under steady external drive, the long-term
behaviour of the fluids is intrinsically unstable: the fluids
never reach a steady state, and rather respond in a unsteady way
(as shown by time measurements of e.g., the shear stress or the
shear rate).
%
%
Such situations include the appearance of \emph{sustained temporal
oscillations} in surfactant
solutions~\cite{Wunenburger,CourbinPanizza} as well as polymer
solutions~\cite{HilliouVlassopoulos}.  In other cases, some
apparently totally erratic temporal responses have been found, in
dense colloidal suspensions~\cite{Laun} and several surfactant
systems~\cite{Bandyopadhyay,Salmon}. In the latter, there are some
indications~\cite{Bandyopadhyay,Salmon} that the obtained signal
is in fact the result of a \emph{deterministic chaotic dynamics}.
Such a chaotic behaviour at virtually zero Reynolds number (and
thus no inertia) must stem from a nonlinearity within the
constitutive properties of the material, and has thus been dubbed
``rheological chaos'', or ``rheochaos''.
\vspace{-.3cm}
\subsubsection{General principles}
\vspace{-.2cm}\enlargethispage*{.5cm}
These unsteady behaviours raise numerous questions from the
theoretical viewpoint. Various hypotheses can be made as to what
mechanisms give rise to such temporal instabilities (oscillations
or rheochaos). The path that we are currently exploring relies on
two main physical features shared by many complex fluids:
\begin{enumerate}[\it (i)]
\item An underlying tendency to form shear-banded flows.
\item A dependence of the present state of the fluid on past history,
due to the presence of slow-relaxing structural modes.
\end{enumerate}

The appearance of shear-bands in shear flows is indeed one of the
most frequent mechanical instability observed in complex fluids.
Slow-relaxing ``structural modes'', on the other hand, are
directly linked to the commonly observed memory effects~-- i.e.,
long-term metastability~-- in complex fluids, where the shear
history of a sample can be remembered for a surprisingly long
time. One way to interpret this is that the previous flow has
disturbed in some way one or several quantities characterizing the
\emph{structure} of the fluid (for instance, the micellar length
in a worm-like micelle system, or the local density in colloidal
systems, etc.), and that these structural variables then only
relax over long periods, because they involve, e.g., collective
motions, or are controlled by a slow, independent physico-chemical
process.

In this article, we give a short overview of our ongoing work on
this subject, presenting two related models which include the
above-mentioned physical ingredients, as well as some preliminary
results.
\vspace{-.3cm}
\subsection{A shear-thickening model with memory}
The first model is a toy rheological equation which describes a
shear-thickening fluid with memory, within a purely scalar
approach where only the shear component $\sigma$ of the stress
tensor is considered. Imagining that the fluid is sheared in a
Couette cell, we take $z$ as the vorticity direction (the axial
direction), and will consider spatial variations in that direction
only. In the proposed model, the evolution with time $t$ of the
shear stress $\sigma(z,t)$ is then given by (in units where the
elastic modulus is one)
\begin{equation}
\label{spaceCHA} \dot\sigma(z,t)= \dot\gamma -R(\sigma) - \lambda
\int_{-\infty}^t\hspace{-.3cm} M(t-t') \,\sigma(z,t')\,
\mathrm{d}t' +\kappa\mbox{\boldmath{$\nabla$}}^2\sigma(z, t)
\end{equation}
with $\dot\sigma\equiv\partial\sigma/\partial t$. The shear rate,
$\dot\gamma$, is supposed \emph{uniform} in the $z$-direction
(this is related to a low-Reynolds assumption). The positive term
$R(\sigma)=a\sigma-b\sigma^2+c\sigma^3$ is a non-linear,
instantaneous relaxation term, with $a$, $b$, $c$ positive
constants chosen so that $R(\sigma)$ has a decreasing portion.
This creates a tendency for the fluid to form ``vorticity'' shear
bands (stacked in the $z$-direction). The integral term over past
states of the stress represents the delayed relaxation of slow
structural modes, with $\lambda>0$ a parameter (homogeneous to a
jump rate) and $M$ a decaying memory function; we will here
specialise to an exponential memory, $M(t)=\taus^{-1}
\exp(-t/\taus)$, and $\taus$ will be the typical relaxation time
of the slow structures. Finally, the inclusion of a non-local
term, in the form of stress diffusion with diffusivity $\kappa$,
is required to describe interfaces between different bands in
inhomogeneous flows.

The original version of this model, proposed by Cates et
al.~\cite{CHA}, was purely temporal (no space variable $z$,
$\kappa=0$) and already proved capable of long-term unsteady
responses: when $\dot\gamma$ is externally fixed in a certain
range, the stress displays sustained (space-homogeneous)
oscillations akin to those of the van der Pol
oscillator\footnote{Regimes of temporal chaos were also found when
the relaxation term $R$ was furthermore allowed to become slightly
delayed. However, the physical meaning of this modification and
its relevance remain uncertain.}. With the spatially-resolved
version presented here, we are able to study the
\emph{spatio-temporal} dynamics of such unstable behaviour.
\vspace{-.5cm}
\subsubsection{Qualitative features of the model}
The qualitative features of the model are best explained by
rewriting the integro-differential equation~\eqref{spaceCHA} as an
equivalent differential system\footnote{This transformation is
possible with an exponential memory, but not in the general
case.}:
\begin{equation}
\label{spaceCHAeq1} \dot\sigma=\dot\gamma-R(\sigma)-\lambda
m+\kappa\nabla^2\sigma\quad,\quad\dot m=-\frac{m-\sigma}{\taus}
\end{equation}
The memory integral within eq.~\eqref{spaceCHA} now appears as an
auxiliary variable $m(z,t)$, which follows its own dynamics and at
each instant tries to relax towards the current value of the
stress $\sigma(z,t)$.

The equilibrium flow curve $\sigma(\dot\gamma)$ for the model is
obtained when $\dot\sigma=\dot m=0$. Flowing states on this curve
are spatially uniform, and on each point of the curve, the memory
$m$ has had time to relax to the equilibrium value of the stress
($m=\sigma$). However, when $\taus$ is large (slow memory
relaxation), this equilibrium flow curve may only be observed on
long enough timescales. On timescales much shorter than $\taus$,
the fluid will behave as if the memory $m$ were static: one has
therefore also to consider a set of ``instantaneous'' flow curves
which correspond to the relaxation of the stress (i.e.,
$\dot\sigma=0$) \emph{at fixed $m$}. As the memory slowly evolves,
the fluid will then accordingly ``jump'' from one instantaneous
flow curve to the other.

%
\begin{figure}
  \includegraphics[width=.45\textwidth]{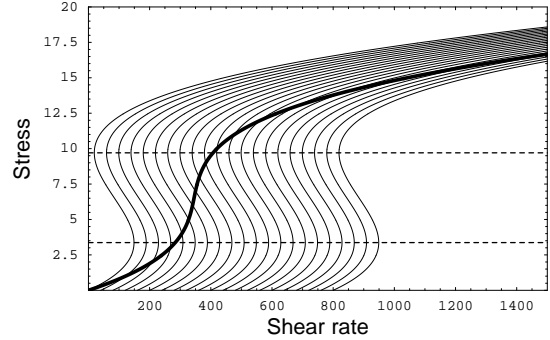}
  \caption{Long-term flow curve (thick line) and underlying short-term curves
  (obtained, from left to right, with fixed values of the memory $m=0$ to
  $m=20$, in steps of one). The stress range between the dotted lines is unstable. (Parameters: $H=1$,
  $\lambda=40$, $\taus=100$, $a=100$, $b=20$, $c=1.02$, $\kappa=0.01$.)}
  \vspace{-.7cm}
\end{figure}
It is then easy to understand qualitatively why the fluid has an
unstable behaviour in a certain range of stress. In Figure~1, the
long-term, equilibrium flow curve is drawn together with the set
of instantaneous curves: the sole inspection of the equilibrium
curve, which has no decreasing portion (for the choice of
parameter values considered), could let one think that all states
are stable; but one observes that there is a region where the
long-term curve is in fact crossed by \emph{decreasing parts} of
short-term curves. Consequently, at these intersection points, the
fluid has an instantaneous tendency to destabilize, thereby
precluding the establishment of the equilibrium state~\cite{CHA}.
\vspace{-.5cm}
\subsubsection{Numerical study of the model}
We now briefly present some of the results that have been found so
far on the spatio-temporal dynamics of the model.

The model was studied numerically through a spectral Galerkin
truncation~\cite{Boyd}, where $\sigma(z,t)$ and $m(z,t)$ are
decomposed as a finite sum of spatial Fourier modes, of the form
$\sigma_n(t) \cos(q_n z)$ with wave-vector $q_n=n \pi/H$ for the
stress ($H$ is the total height  of our imaginary Couette device),
plus a similar set of memory modes $m_n(t)$. (We usually take the
first ten modes into account in our numerics.) The system of
equations~\eqref{spaceCHAeq1} then reduces to a set of coupled
ordinary differential equations governing the temporal evolution
of $\sigma_n(t)$ and $m_n(t)$.

As is the case for conventional shear bands, two different
protocols can be followed in the numerical study of the
instability (details will appear elsewhere~\cite{Aradian}): either
working at fixed shear rate $\dot\gamma$ (i.e., with a
shear-controlled Couette cell), or working at fixed torque, or
equivalently at fixed average stress $\langle \sigma \rangle$
(stress-controlled Couette). These two protocols lead to rather
different results, as we shall now see.
%
\subsubsection{Working at fixed $\dot\gamma$}
We here work with a fixed, externally imposed value of
$\dot\gamma$, and compute the values of the different modes
$\sigma_n(t)$ and $m_n(t)$, from which we reconstruct
$\sigma(z,t)$ and $m(z,t)$. For numerical solutions carried within
the unstable window, our results show that, in the vast majority
of cases, the spatio-temporal dynamics of the instability is in
fact essentially purely temporal: regardless of their initial
magnitude, all the modes rapidly decay and disappear, except for
the spatially uniform mode $\sigma_0(t)$. This mode then
oscillates alone, with a fixed period and a regular shape, as in
the temporal version of ref.~\cite{CHA}.

In a few, small regions of the parameter space explored, it has
been possible to obtain a slightly richer dynamics, where some of
the lower modes survive and undergo small-scale periodic
oscillations, alongside a large uniform mode oscillation which
still dominates.
%
%
\subsubsection{Working at fixed $\langle\sigma\rangle$}
We now work at imposed average stress $\langle\sigma\rangle$; in
terms of Fourier modes, this corresponds to fixing the value of
the uniform Fourier mode $\sigma_0(t)=\mbox{const.}$, which will
thus not be able to oscillate at all. A very rich spatio-temporal
behaviour then arises, as the higher spatial modes will now, by
necessity, be involved in the instability.

Different unstable features may appear~\cite{Aradian}; In
Figure~2-\emph{a}, we present a most striking one, called
``flip-flop shear-bands'': a shear-banded profile appears, with a
low shear band and a high shear band separated by an interface;
but these bands are unstable, and periodically flip, the higher
band becoming the lower and vice-versa. The ``flipping time'' is
very short as compared to the latency time between two flips.
Figure~2-\emph{b} shows the corresponding temporal evolution of
the total stress $\sigma(z,t)$.
\enlargethispage*{.5cm}

Much more irregular-looking time variations can also be obtained
for different parameter choices, as Fig.~2-\emph{c} shows for the
stress. We emphasize however that these patterns remain
time-periodic; so far no chaotic response has been found in the
present model.
\begin{figure}
  \includegraphics[clip=true,bb=3.3cm 12.7cm 16.6cm 24.3cm,width=.45\textwidth]{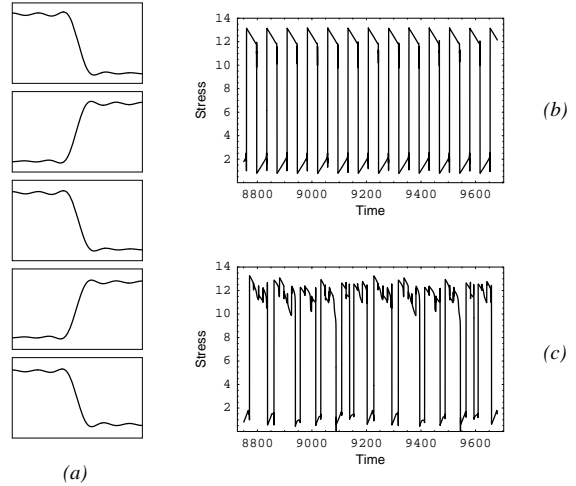}
  \caption{\emph{(a)}~``Flip-flop'' shear-bands for $\langle\sigma\rangle=7.0$: From top to bottom,
  successive snapshots of the stress $\sigma(z,t)$ vs. position $z$, at times $t=8800,8840,8880,
  8920,8960$. \emph{(b)}~Time series of the stress at position $z=0$ for the same value of $\langle\sigma\rangle$.
  \emph{(c)}~Time series of the stress at position $z=0$ for $\langle\sigma\rangle=9.0$. The
  time window corresponds to two periods of the signal. (Parameters same as Fig.~1.)
  \vspace{-.3cm}}
\end{figure}
\vspace{-.3cm} \enlargethispage*{.4cm}
\subsection{A model with ``fluidity''}
We would now like to introduce another fluid model, based on the
same general ideas as previously, but more specifically oriented
towards solutions of wormlike micelles and polymers. The shear
stress $\Sigma$ in the fluid is locally the sum of a polymer or
micellar part, $\sigma$, and a Newtonian part $\eta\dot\gamma$
corresponding to the solvent:
\begin{equation}
\label{totstress}
\Sigma=\sigma + \eta\dot\gamma
\end{equation}
The model's equations are as follows:
\begin{eqnarray}
\label{sigmaeq}
\dot\sigma&=&-\frac{\sigma}{\tau}+\frac{G(\tau)}{\tau}g(\dot\gamma\tau)+\kappa\nabla^2\sigma\\
\label{fluidityeq}
\dot\tau&=&-\frac{\tau-\taueq(\dot\gamma)}{\taus}
\end{eqnarray}

The stress evolution equation~\eqref{sigmaeq} has a classical
form, and describes the relaxation (with a timescale $\tau$) of
the ``polymer'' stress towards a value which, at a given
$\dot\gamma$, is controlled by $ G(\tau) g(\dot\gamma\tau)$.
$G(\tau)$ is the elastic modulus and generally depends on the
value of $\tau$ (see below). The function $g$ is hump-shaped,
hence conferring on the fluid a tendency to shear-thinning, and
shear-banding, with bands in the velocity gradient direction
(i.e., the radial direction of the Couette cell).

Similarly to the  model with memory, one of the variables involved
in the stress equation is subject to a distinct dynamics: this is
now the Maxwell time $\tau$ of the fluid, which, as stated by
eq.~\eqref{fluidityeq}, relaxes towards a shear-rate dependent
equilibrium value $\taueq(\dot\gamma)$ with a characteristic time
$\taus$. We note that having a dynamical Maxwell time is indeed
very similar to the ``fluidity'' model introduced by Derec et al.\
in the context of paste flow (see~\cite{Derec}).

Here again, $\tau$ is a ``structural'' variable in the sense that
it reflect variations in the local structure of the fluid: for
semi-dilute solutions of wormlike micelles, it will typically
relate to local variations of the mean chain length of the
micelles. In this interpretation, the chain length distribution,
and consequently the mean chain length, may change only through
the action of some micelle-micelle chemical reactions, which will
have their own timescale $\taus$~-- this time will thus control
the relaxation of $\tau$, as in eq.~\eqref{fluidityeq}. On the
other hand, the dependence of the equilibrium value $\taueq$ on
the shear rate can be interpreted as a displacement of the
chemical equilibrium by the mechanical shearing (for example,
decreasing $\taueq$, by helping chain scission, or, increasing
$\taueq$, by helping polymerisation). For more dilute solutions
(non-entangled), one may rather interpret the variations of
$\taueq$ as related to the strong shear-thickening transition
usually observed upon varying $\dot\gamma$: the structural changes
occurring at the transition (gelation) will then strongly alter
$\taueq$.

%
In accord with the model with memory of the previous section, we
have focussed on slow relaxations of the structural variable:
$\taus \gg 1$. Then one can again construct a ``long-term'',
master flow curve where all variables are equilibrated
($\dot\sigma=\dot\tau=0$), and an underlying set of ``short-term''
curves where $\tau$ is fixed (quasi-static) and the stress is
equilibrated with respect to that value of $\tau$. As seen in the
previous model, an increasing portion of the ``long-term'' curve
is here also made unstable when crossed by decreasing portions of
underlying curves.

Two variants of the model can be studied, depending on whether the
elastic modulus $G$ is affected or not by changes in the structure
of the fluid, that is, changes in $\tau$. Variant~1 corresponds to
$G(\tau)=G_0$ being constant, which would be suitable for a
semi-dilute, entangled solution of micelles, where (reasonable)
changes in the mean micellar length leave the modulus unchanged.
In variant~2, $G$ varies with $\tau$, as would
 e.g.\ be the case in the shear-thickening transition of dilute micelles,
where the significant structural changes affecting $\tau$ will
affect $G$ as well.
\begin{figure}
  \includegraphics[width=.40\textwidth,clip=true,bb=6.3cm 9.5cm 14.5cm 17.5cm]{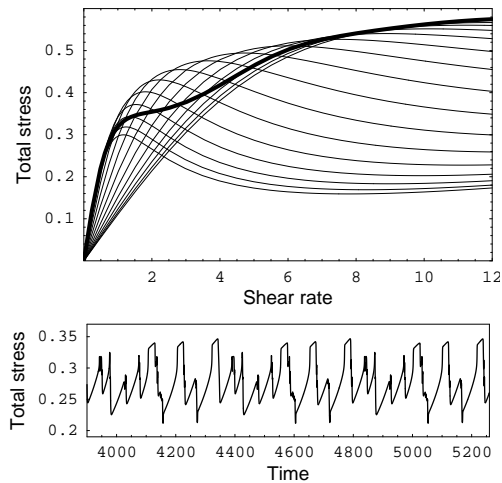}
  \caption{Upper figure: Long-term flow curve (thick line) and underlying short-term curves
  in variant~2. Lower figure: very irregular (though periodic) time series of the total
  stress $\Sigma(t)$ for $\langle\dot\gamma\rangle=4.0$.
  \vspace{-.5cm}}
\end{figure}
Work on both variants of the model is currently in progress, but
results so far are promising. Figure~3 shows an example of results
in variant~2.
%

To conclude, we would like to mention extremely interesting
results by S.~Fielding and P.~Olmsted~\cite{Fielding}: working
independently on essentially the same model as
eqs.~\eqref{sigmaeq}-\eqref{fluidityeq}, they have been able,
within variant~1, to obtain chaotic-like signals in a regime where
\emph{both} the underlying curves and the master flow curve have
decreasing portions, \emph{and} where the structural relaxation
occurs on timescales comparable to the stress relaxation
($\taus\simeq 1$).

Many questions are still open in the study of the type of models
which have been described in this article. Are the existence of
distinct ``structural'' relaxations really the origin of the
instabilities observed in experiments? Is the regime just
described the only one displaying chaotic-like behaviour, and
relatedly, what is required for regular or irregular-looking
periodic motion to destabilize into chaos?
%
%
%
%
\vspace{-.5cm}
\begin{theacknowledgments}
The authors would like to thank S.~Fielding and P.~Olmsted for
sharing their work with them prior to publication. AA is funded by
an EPSRC Post-doctoral Fellowship in Theoretical Physics
(GR/R95098).
\end{theacknowledgments}
%
%

%
%
%
%
%
\end{document}